\PassOptionsToPackage{unicode}{hyperref}
\PassOptionsToPackage{hyphens}{url}
\PassOptionsToPackage{dvipsnames,svgnames,x11names}{xcolor}
\documentclass[
]{article}
\usepackage{amsmath,amssymb}
\usepackage{lmodern}
\usepackage{iftex}
\ifPDFTeX
  \usepackage[T1]{fontenc}
  \usepackage[utf8]{inputenc}
  \usepackage{textcomp} 
\else 
  \usepackage{unicode-math}
  \defaultfontfeatures{Scale=MatchLowercase}
  \defaultfontfeatures[\rmfamily]{Ligatures=TeX,Scale=1}
\fi
\IfFileExists{upquote.sty}{\usepackage{upquote}}{}
\IfFileExists{microtype.sty}{
  \usepackage[]{microtype}
  \UseMicrotypeSet[protrusion]{basicmath} 
}{}
\makeatletter
\@ifundefined{KOMAClassName}{
  \IfFileExists{parskip.sty}{%
    \usepackage{parskip}
  }{
    \setlength{\parindent}{0pt}
    \setlength{\parskip}{6pt plus 2pt minus 1pt}}
}{
  \KOMAoptions{parskip=half}}
\makeatother
\usepackage{xcolor}
\usepackage{color}
\usepackage{fancyvrb}

\DefineVerbatimEnvironment{Highlighting}{Verbatim}{commandchars=\\\{\}}
\newenvironment{Shaded}{}{}

\newcommand{\AttributeTok}[1]{\textcolor[rgb]{0.49,0.56,0.16}{#1}}

\newcommand{\DecValTok}[1]{\textcolor[rgb]{0.25,0.63,0.44}{#1}}

\newcommand{\ExtensionTok}[1]{#1}

\newcommand{\FunctionTok}[1]{\textcolor[rgb]{0.02,0.16,0.49}{#1}}

\newcommand{\KeywordTok}[1]{\textcolor[rgb]{0.00,0.44,0.13}{\textbf{#1}}}
\newcommand{\NormalTok}[1]{#1}

\usepackage{graphicx}
\makeatletter
\def\maxwidth{\ifdim\Gin@nat@width>\linewidth\linewidth\else\Gin@nat@width\fi}
\def\maxheight{\ifdim\Gin@nat@height>\textheight\textheight\else\Gin@nat@height\fi}
\makeatother
\setkeys{Gin}{width=\maxwidth,height=\maxheight,keepaspectratio}
\makeatletter
\def\fps@figure{htbp}
\makeatother
\NewDocumentCommand\citeproctext{}{}
\NewDocumentCommand\citeproc{mm}{%
  \begingroup\def\citeproctext{#2}\cite{#1}\endgroup}
\makeatletter
 \let\@cite@ofmt\@firstofone
 \def\@biblabel#1{}
 \def\@cite#1#2{{#1\if@tempswa , #2\fi}}
\makeatother
\newlength{\cslhangindent}
\newlength{\csllabelwidth}
\newenvironment{CSLReferences}[2] 
 {\begin{list}{}{%
  \setlength{\itemindent}{0pt}
  \setlength{\leftmargin}{0pt}
  \setlength{\parsep}{0pt}
  \ifodd #1
   \setlength{\leftmargin}{\cslhangindent}
   \setlength{\itemindent}{-1\cslhangindent}
  \fi
  \setlength{\itemsep}{#2\baselineskip}}}
 {\end{list}}
\usepackage{calc}

\ifLuaTeX
\usepackage[bidi=basic]{babel}
\else
\usepackage[bidi=default]{babel}
\fi
\babelprovide[main,import]{american}

\def\languageshorthands#1{}
\ifLuaTeX
  \usepackage{selnolig}  
\fi
\IfFileExists{bookmark.sty}{\usepackage{bookmark}}{\usepackage{hyperref}}
\IfFileExists{xurl.sty}{\usepackage{xurl}}{} 
\hypersetup{
  pdftitle={Metewand: Reproducible Computational Benchmarks Across
Languages},
  pdfauthor={Johan Larsson},
  pdflang={en-US},
  colorlinks=true,
  linkcolor={Maroon},
  filecolor={Maroon},
  citecolor={Blue},
  urlcolor={Blue},
  pdfcreator={LaTeX via pandoc}}

\title{Metewand: Reproducible Computational Benchmarks Across Languages}

\definecolor{c53baa1}{RGB}{83,186,161}
\definecolor{c202826}{RGB}{32,40,38}
\def \rorglobalscale {0.1}
\newcommand{\rorlogo}{%
\begin{tikzpicture}[y=1cm, x=1cm, yscale=\rorglobalscale,xscale=\rorglobalscale, every node/.append style={scale=\rorglobalscale}, inner sep=0pt, outer sep=0pt]
  \begin{scope}[even odd rule,line join=round,miter limit=2.0,shift={(-0.025, 0.0216)}]
    \path[fill=c53baa1,nonzero rule,line join=round,miter limit=2.0] (1.8164, 3.012) -- (1.4954, 2.5204) -- (1.1742, 3.012) -- (1.8164, 3.012) -- cycle;
    \path[fill=c53baa1,nonzero rule,line join=round,miter limit=2.0] (3.1594, 3.012) -- (2.8385, 2.5204) -- (2.5172, 3.012) -- (3.1594, 3.012) -- cycle;
    \path[fill=c53baa1,nonzero rule,line join=round,miter limit=2.0] (1.1742, 0.0669) -- (1.4954, 0.5588) -- (1.8164, 0.0669) -- (1.1742, 0.0669) -- cycle;
    \path[fill=c53baa1,nonzero rule,line join=round,miter limit=2.0] (2.5172, 0.0669) -- (2.8385, 0.5588) -- (3.1594, 0.0669) -- (2.5172, 0.0669) -- cycle;
    \path[fill=c202826,nonzero rule,line join=round,miter limit=2.0] (3.8505, 1.4364).. controls (3.9643, 1.4576) and (4.0508, 1.5081) .. (4.1098, 1.5878).. controls (4.169, 1.6674) and (4.1984, 1.7642) .. (4.1984, 1.8777).. controls (4.1984, 1.9719) and (4.182, 2.0503) .. (4.1495, 2.1132).. controls (4.1169, 2.1762) and (4.0727, 2.2262) .. (4.0174, 2.2635).. controls (3.9621, 2.3006) and (3.8976, 2.3273) .. (3.824, 2.3432).. controls (3.7505, 2.359) and (3.6727, 2.367) .. (3.5909, 2.367) -- (2.9676, 2.367) -- (2.9676, 1.8688).. controls (2.9625, 1.8833) and (2.9572, 1.8976) .. (2.9514, 1.9119).. controls (2.9083, 2.0164) and (2.848, 2.1056) .. (2.7705, 2.1791).. controls (2.6929, 2.2527) and (2.6014, 2.3093) .. (2.495, 2.3487).. controls (2.3889, 2.3881) and (2.2728, 2.408) .. (2.1468, 2.408).. controls (2.0209, 2.408) and (1.905, 2.3881) .. (1.7986, 2.3487).. controls (1.6925, 2.3093) and (1.6007, 2.2527) .. (1.5232, 2.1791).. controls (1.4539, 2.1132) and (1.3983, 2.0346) .. (1.3565, 1.9436).. controls (1.3504, 2.009) and (1.3351, 2.0656) .. (1.3105, 2.1132).. controls (1.2779, 2.1762) and (1.2338, 2.2262) .. (1.1785, 2.2635).. controls (1.1232, 2.3006) and (1.0586, 2.3273) .. (0.985, 2.3432).. controls (0.9115, 2.359) and (0.8337, 2.367) .. (0.7519, 2.367) -- (0.1289, 2.367) -- (0.1289, 0.7562) -- (0.4837, 0.7562) -- (0.4837, 1.4002) -- (0.6588, 1.4002) -- (0.9956, 0.7562) -- (1.4211, 0.7562) -- (1.0118, 1.4364).. controls (1.1255, 1.4576) and (1.2121, 1.5081) .. (1.2711, 1.5878).. controls (1.2737, 1.5915) and (1.2761, 1.5954) .. (1.2787, 1.5991).. controls (1.2782, 1.5867) and (1.2779, 1.5743) .. (1.2779, 1.5616).. controls (1.2779, 1.4327) and (1.2996, 1.3158) .. (1.3428, 1.2113).. controls (1.3859, 1.1068) and (1.4462, 1.0176) .. (1.5237, 0.944).. controls (1.601, 0.8705) and (1.6928, 0.8139) .. (1.7992, 0.7744).. controls (1.9053, 0.735) and (2.0214, 0.7152) .. (2.1474, 0.7152).. controls (2.2733, 0.7152) and (2.3892, 0.735) .. (2.4956, 0.7744).. controls (2.6016, 0.8139) and (2.6935, 0.8705) .. (2.771, 0.944).. controls (2.8482, 1.0176) and (2.9086, 1.1068) .. (2.952, 1.2113).. controls (2.9578, 1.2253) and (2.9631, 1.2398) .. (2.9681, 1.2544) -- (2.9681, 0.7562) -- (3.3229, 0.7562) -- (3.3229, 1.4002) -- (3.4981, 1.4002) -- (3.8349, 0.7562) -- (4.2603, 0.7562) -- (3.8505, 1.4364) -- cycle(0.9628, 1.7777).. controls (0.9438, 1.7534) and (0.92, 1.7357) .. (0.8911, 1.7243).. controls (0.8623, 1.7129) and (0.83, 1.706) .. (0.7945, 1.7039).. controls (0.7588, 1.7015) and (0.7252, 1.7005) .. (0.6932, 1.7005) -- (0.4839, 1.7005) -- (0.4839, 2.0667) -- (0.716, 2.0667).. controls (0.7477, 2.0667) and (0.7805, 2.0643) .. (0.8139, 2.0598).. controls (0.8472, 2.0553) and (0.8768, 2.0466) .. (0.9025, 2.0336).. controls (0.9282, 2.0206) and (0.9496, 2.0021) .. (0.9663, 1.9778).. controls (0.9829, 1.9534) and (0.9914, 1.9209) .. (0.9914, 1.8799).. controls (0.9914, 1.8362) and (0.9819, 1.8021) .. (0.9628, 1.7777) -- cycle(2.6125, 1.3533).. controls (2.5889, 1.2904) and (2.5553, 1.2359) .. (2.5112, 1.1896).. controls (2.4672, 1.1433) and (2.4146, 1.1073) .. (2.3529, 1.0814).. controls (2.2916, 1.0554) and (2.2228, 1.0427) .. (2.1471, 1.0427).. controls (2.0712, 1.0427) and (2.0026, 1.0557) .. (1.9412, 1.0814).. controls (1.8799, 1.107) and (1.8272, 1.1433) .. (1.783, 1.1896).. controls (1.7391, 1.2359) and (1.7052, 1.2904) .. (1.6817, 1.3533).. controls (1.6581, 1.4163) and (1.6465, 1.4856) .. (1.6465, 1.5616).. controls (1.6465, 1.6359) and (1.6581, 1.705) .. (1.6817, 1.7687).. controls (1.7052, 1.8325) and (1.7388, 1.8873) .. (1.783, 1.9336).. controls (1.8269, 1.9799) and (1.8796, 2.0159) .. (1.9412, 2.0418).. controls (2.0026, 2.0675) and (2.0712, 2.0804) .. (2.1471, 2.0804).. controls (2.223, 2.0804) and (2.2916, 2.0675) .. (2.3529, 2.0418).. controls (2.4143, 2.0161) and (2.467, 1.9799) .. (2.5112, 1.9336).. controls (2.5551, 1.8873) and (2.5889, 1.8322) .. (2.6125, 1.7687).. controls (2.636, 1.705) and (2.6477, 1.6359) .. (2.6477, 1.5616).. controls (2.6477, 1.4856) and (2.636, 1.4163) .. (2.6125, 1.3533) -- cycle(3.8015, 1.7777).. controls (3.7825, 1.7534) and (3.7587, 1.7357) .. (3.7298, 1.7243).. controls (3.701, 1.7129) and (3.6687, 1.706) .. (3.6333, 1.7039).. controls (3.5975, 1.7015) and (3.5639, 1.7005) .. (3.5319, 1.7005) -- (3.3226, 1.7005) -- (3.3226, 2.0667) -- (3.5547, 2.0667).. controls (3.5864, 2.0667) and (3.6192, 2.0643) .. (3.6526, 2.0598).. controls (3.6859, 2.0553) and (3.7155, 2.0466) .. (3.7412, 2.0336).. controls (3.7669, 2.0206) and (3.7883, 2.0021) .. (3.805, 1.9778).. controls (3.8216, 1.9534) and (3.8301, 1.9209) .. (3.8301, 1.8799).. controls (3.8301, 1.8362) and (3.8206, 1.8021) .. (3.8015, 1.7777) -- cycle;
  \end{scope}
\end{tikzpicture}
}

\usepackage[affil-it]{authblk}
\usepackage{orcidlink}
\author[1%
  ]{Johan Larsson%
    \,\orcidlink{0000-0002-4029-5945}\,%
    }

\affil[1]{Department of Mathematical Sciences, University of Copenhagen,
Denmark%
    \,\protect\href{https://ror.org/035b05819}{\protect\rorlogo}\,%
  }
\date{30 November 2026}

\begin{document}
\maketitle

\section{Summary}\label{summary}

\href{https://github.com/jolars/metewand}{Metewand} is a framework and
command-line tool for benchmarking computational methods across
programming languages. To set up a benchmark, the researcher specifies
which methods to run on which data and supplies an independent evaluator
that checks the answers. Metewand then runs each method in a separate
process and records how long it took, how well its answer met the
evaluation criteria, and whether the run failed. It also records which
data, code, and settings produced each measurement. This makes it
possible to compare both speed and accuracy in problems such as
numerical optimization and statistical estimation. The command-line tool
runs on Linux, macOS, and Windows. The project also includes small
Python and R packages that make it easier to write benchmarks in those
languages. Metewand is open source under either the MIT License or the
Apache License, Version 2.0. Its documentation is available at
\href{https://metewand.org}{metewand.org}.

\section{Statement of Need}\label{statement-of-need}

Benchmarks help researchers decide which algorithms to develop and which
software to use. But for a comparison to be meaningful, the methods
studied must use the same inputs, solve the same problem, and be judged
by the same accuracy criteria. For example, one solver may appear faster
because it stops at a looser tolerance. Another may fail on some of the
problems in the benchmark. Both outcomes matter when constructing a
performance profile~(\citeproc{ref-dolan2002}{Dolan \& Moré, 2002}),
which relies on measurements defined consistently across problems.

These may seem like simple requirements, but the devil is in the
details. A benchmark script often generates data, calls solvers
(potentially written in different languages), measures their runtimes,
checks their answers, and summarizes the results. Each of these steps
can matter for the results and for reproducibility. For example, a data
generator may use a random seed that is not recorded, so the benchmark
cannot be repeated. A solver may require a specific version of a
library, and the benchmark may not record which version was used.

In Metewand, the problem defines what the methods should compute, what
makes an answer valid, and what target, if any, it must reach. An
independent evaluator applies the same rules to every implementation.
The implementations share the same data, which the benchmark can load or
generate. Problems that need no external data can omit the dataset. The
author chooses which parts of each run to time, and Metewand records the
inputs and settings actually used, the timing policy, and the outcomes.

\section{State of the Field}\label{state-of-the-field}

hyperfine runs commands repeatedly and provides warm-ups and statistical
summaries of runtime~(\citeproc{ref-hyperfine}{Peter, n.d.}). airspeed
velocity tracks runtime, memory use, and custom measurements across
versions of a Python package~(\citeproc{ref-asv}{asv Developers, n.d.}).
IOHprofiler supports experiments with iterative optimization heuristics
and provides tools for comparing them at fixed targets and fixed
budgets~(\citeproc{ref-doerr2018}{Doerr et al., 2018}).

The closest relative to Metewand is Benchopt, a Python framework that
organizes optimization benchmarks around datasets, objectives, and
solvers. Benchopt supports algorithms written in several languages and
can install benchmark dependencies with
Conda~(\citeproc{ref-moreau2022}{Moreau et al., 2022}). Metewand makes a
similar separation between the problem and the methods used to solve it.
It runs data generators, methods, and evaluators as separate processes,
all using the same versioned protocol. Benchmark authors can choose the
language and software environment for each worker. For example, they can
use renv to set up a local R environment.

Unlike Benchopt, Metewand is not intended to be a collaborative platform
for benchmarks. There is no official collection of benchmarks and no
public repository of benchmark results that users can contribute to.

\section{Software Design}\label{software-design}

A benchmark starts with a YAML file called a manifest. It says which
data to use, what problem to solve, and which methods to compare. For
example, the sample size is a setting for a simulated dataset, an
accuracy target belongs to the problem, and an iteration limit belongs
to the method. The manifest also says what to time and how to record
progress. Metewand turns grids of parameter values and named cases into
a plan that the author can inspect before running the benchmark. Figure
1 follows a least-squares comparison from manifest to measurements.

\begin{figure}
\centering
\includegraphics[width=0.96\textwidth,height=\textheight]{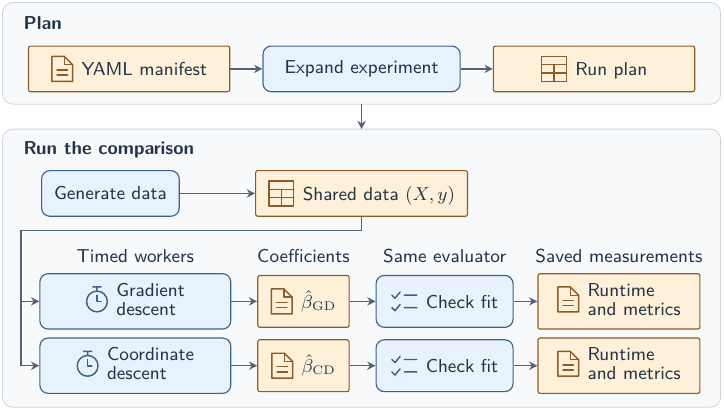}
\caption{The least-squares workflow for two successful runs. Gold boxes
with file or table icons show inputs and outputs. Rounded blue boxes
show actions. Each stopwatch marks a separately timed worker. The same
evaluator computes each fit's objective value and gradient norm from the
shared data.}
\end{figure}

A compiled program and an R script can take part in the same benchmark.
Each runs in its own process, called a worker, and talks to Metewand
through the same versioned protocol. Workers write their results to
files that Metewand checks. The Python and R packages let authors write
ordinary functions to prepare a method's inputs or solve the problem.
Metewand's worker adapters handle the protocol for those functions.

Metewand checks that a method's answer has the structure the problem
requires. The evaluator computes metrics and applies the problem's rules
for a valid answer. For example, it might reject a solution that
violates a constraint. The evaluator also checks whether the answer
meets any accuracy target set by the problem. Metewand keeps the metrics
for valid answers even when they miss the target. When there is no
target, a successful run means the answer passed validation. Authors can
test the evaluator on cases with known answers before running the
benchmark.

By default, Metewand measures the wall time spent preparing the method,
running it, and writing its answer. Metewand builds the shared dataset
and starts the worker before timing the method, then evaluates the
answer afterward. If the problem allows it, the author can include
worker startup or start the timer after preparation. A worker can also
time its solver function. Metewand records that time separately, and
some policies can use it as the primary measurement.

To follow a solver's progress, the author can run it from scratch at
different budgets, for example by raising the iteration limit for each
run. A solver that saves intermediate answers can also provide snapshots
during a single run. The evaluator checks each answer independently, and
Metewand records how the answers were collected and which clocks were
used.

Metewand gives benchmark settings and the inputs it actually uses stable
identifiers by hashing a standard representation of each. It derives
separate seeds for the data generator, the methods, and the scheduler.
The generator and methods can use their seeds to initialize their own
random number generators. Methods in the same comparison receive the
same seed for a given repetition. Metewand can use the scheduling seed
to shuffle their order reproducibly.

When a benchmark stops partway through, Metewand can resume it using the
measurements it has already accepted. If the author retries a failure,
Metewand adds a new attempt to the history. It records attempts
separately from measurements based on valid answers, so researchers can
see when a method failed during execution, missed the target, or yielded
an accepted measurement. For local runs, Metewand uses SQLite
transactions to keep the records consistent and saves result files as
immutable artifacts. An archive lets others inspect and export the
results offline without the original checkout.

Before starting workers, Metewand can verify prepared Nix or uv
environments against the evidence in its lockfile. On Linux, it can also
check Docker OCI or Apptainer images against the lockfile before running
workers in them. The lockfile records details such as the digest of a
Docker image. Workers may access the filesystem and network, so Metewand
assumes their code can be trusted. The recorded settings and software
help researchers recreate the setup. Timings can still vary with the
hardware and what else the machine is doing.

\section{Example}\label{example}

The least-squares example compares gradient descent with cyclic
coordinate descent on a simulated dataset. The manifest defines the data
generator, the two methods, and the problem's evaluator. The experiment
below brings them together and sets the random seed.

\begin{Shaded}
\begin{Highlighting}[]
\FunctionTok{experiments}\KeywordTok{:}
\AttributeTok{  }\KeywordTok{{-}}\AttributeTok{ }\FunctionTok{name}\KeywordTok{:}\AttributeTok{ comparison}
\AttributeTok{    }\FunctionTok{problem}\KeywordTok{:}\AttributeTok{ least\_squares}
\AttributeTok{    }\FunctionTok{datasets}\KeywordTok{:}\AttributeTok{ }\KeywordTok{[}\AttributeTok{simulated}\KeywordTok{]}
\AttributeTok{    }\FunctionTok{implementations}\KeywordTok{:}\AttributeTok{ }\KeywordTok{[}\AttributeTok{gradient\_descent}\KeywordTok{,}\AttributeTok{ coordinate\_descent}\KeywordTok{]}
\AttributeTok{    }\FunctionTok{seed}\KeywordTok{:}\AttributeTok{ }\DecValTok{7}
\end{Highlighting}
\end{Shaded}

The evaluator uses the coefficients from each method to compute the
objective value and the gradient norm. A second manifest sets a target
for the gradient norm and supplies cases with known answers to check the
evaluator. It also runs each method for just one iteration. Both methods
return valid coefficients in those runs, and their gradient norms are
still above the target.

With the Metewand Python package installed, run these commands from the
example directory to preview the benchmark, run it, inspect the results,
and export them as CSV.

\begin{Shaded}
\begin{Highlighting}[]
\ExtensionTok{metewand}\NormalTok{ plan}
\ExtensionTok{metewand}\NormalTok{ run}
\ExtensionTok{metewand}\NormalTok{ status}
\ExtensionTok{metewand}\NormalTok{ export }\AttributeTok{{-}{-}output{-}dir}\NormalTok{ .metewand }\AttributeTok{{-}{-}format}\NormalTok{ csv }\AttributeTok{{-}{-}destination}\NormalTok{ tables}
\end{Highlighting}
\end{Shaded}

Omit \texttt{-\/-format\ csv} to export JSONL. The manifests are in the
\href{https://github.com/jolars/metewand/tree/main/examples/least-squares}{least-squares
example}.

A third manifest, \texttt{profile.yaml}, follows one solve per method on
the same seeded dataset. At selected iterations, Metewand pauses the
worker to evaluate its current coefficients and stops the run once the
gradient norm meets the target. In Figure 2, cyclic coordinate descent
has an earlier checkpoint at or below the target on this dataset.

\begin{figure}
\centering
\includegraphics[width=0.96\textwidth,height=\textheight]{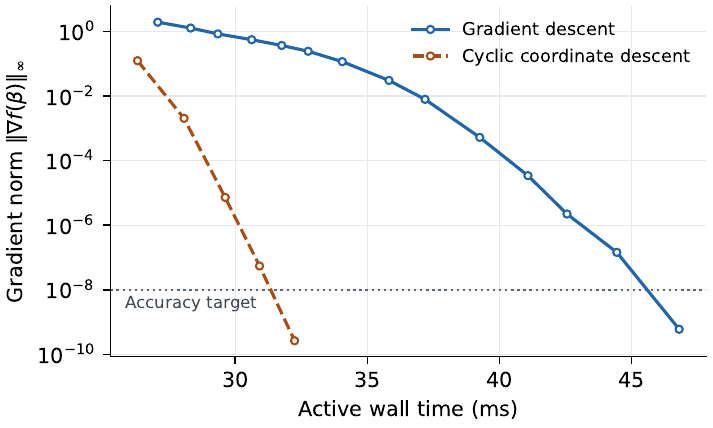}
\caption{Gradient norm along single solves on the simulated
least-squares dataset. Circles mark independently evaluated checkpoints,
and the dotted line marks the target of \(10^{-8}\). Active wall time
includes writing checkpoints and excludes pauses for evaluation.}
\end{figure}

\section{AI Usage Disclosure}\label{ai-usage-disclosure}

OpenAI Codex helped edit this manuscript, assemble the bibliography, and
create the figures.

\section{Acknowledgements}\label{acknowledgements}

The inspiration for Metewand came from being a part of the Benchopt
project, and I am grateful to the Benchopt team for their work and for
sharing their experiences.

\section*{References}\label{references}
\addcontentsline{toc}{section}{References}

\phantomsection\label{refs}

\end{document}